\documentclass[twocolumn,showpacs,amsmath,amssymb,amsbsy,superscriptaddress,prl]{revtex4-1}
\pdfoutput=1

\usepackage[dvipdfmx]{graphicx}
\usepackage{rotating}
\usepackage{dcolumn}
\usepackage{bm}
\usepackage{amsfonts}
\usepackage[abs]{overpic}
\usepackage{color}
\usepackage{url}
\usepackage{accents}
\newcommand{\angstrom}{\textup{\AA}}

\begin{document}


\title{Electron-phonon coupling mechanisms for hydrogen-rich metals at high pressure}

\author{K. Tanaka}
\email{kat221@campus.usask.ca}
\affiliation{Department of Physics and Engineering Physics, University of Saskatchewan, 116 Science Place, Saskatoon, SK, S7N 5E2 Canada}

\author{J. S. Tse}
\email{john.tse@usask.ca}
\affiliation{Department of Physics and Engineering Physics, University of Saskatchewan, 116 Science Place, Saskatoon, SK, S7N 5E2 Canada}
\affiliation{State Key Laboratory for Superhard Materials, Jilin University, Changchun 130012, China}

\author{H. Liu}
\affiliation{Department of Physics and Engineering Physics, University of Saskatchewan, 116 Science Place, Saskatoon, SK, S7N 5E2 Canada}
\affiliation{Geophysical Laboratory, Carnegie Institution for Science, Washington, DC 20015, USA}

\date{\today}
             
\begin{abstract}
The mechanisms for strong electron-phonon coupling predicted for hydrogen-rich alloys with high superconducting critical temperature ($T_c$) are examined within the Migdal-Eliashberg theory. Analysis of the functional derivative of $T_c$ with respect to the electron-phonon spectral function shows that at low pressures, when the alloys often adopt layered structures, bending vibrations have the most dominant effect. At very high pressures, the H-H interactions in two-dimensional (2D) and three-dimensional (3D) extended structures are weakened, resulting in mixed bent (libration) and stretch vibrations, and the electron-phonon coupling process is distributed over a broad frequency range leading to very high $T_c$.
\end{abstract}

\pacs{
74.20.Pq, 	
74.25.Kc, 	
74.62.Bf, 	
74.62.Fj, 	
}
\maketitle

Hydrogen is the lightest element and thus, if molecular hydrogen can be compressed into a metal in the solid state, it is expected to become superconducting at a very high transition temperature $T_c$ due to exceptionally strong electron-phonon coupling \cite{Ashcroft1968}. Although this proposal has not been verified experimentally, calculation based on modern electronic structure theory has predicted high $T_c$ for metallic hydrogen \cite{Cudazzo2010,McMahon2011}.  According to recent studies, metallization of solid hydrogen may require pressure in excess of 400 GPa, which is difficult to achieve with today's experimental techniques. Although very recently metallic hydrogen has been claimed to be observed under extremely high pressure of nominal 495 GPa \cite{Dias2017}, the result is contentious \cite{metallicH} and further experiments are required. On the other hand,
it has been realised that the electron density required to metallize molecular hydrogen may be achieved by compression of group-IVa hydrides, in which the hydrogen content is already high \cite{Ashcroft2004}. This insightful suggestion has stimulated various theoretical studies and experimental investigations. The first prediction of superconductivity in group-IVa hydrides was made using density functional theory for silane (SiH$_4$) \cite{Feng2006}. First-principles calculation has predicted a monoclinic $C2/c$ metallic solid composed of SiH$_4$ layers bridged by Si-H-Si bonds that is stable between 65 and 150 GPa and has $T_c$ of 45-55 K at 90-125 GPa \cite{Yao2008}. Subsequent experiment has revealed that insulating molecular silane transforms to a metal at 50 GPa and becomes superconducting with $T_c=17$ K at 96 and 120 GPa \cite{Eremets2008}. However, the measured diffraction pattern of the superconducting phase did not match the monoclinic structure \cite{Eremets2008} and the high-pressure structure of silane remains controversial \cite{Strobel2011}. 

In the past decade, numerous theoretical predictions have been made on the structure and superconducting behaviour of stoichiometric and hydrogen-enriched hydrides with a variety of elements at high pressures. Most notable examples are the high-pressure polymorphs of CaH$_6$ and YH$_6$, both of which have a novel cagelike structure formed by monatomic hydrogens and have been predicted to have $T_c$ higher than 200 K \cite{Wang2012,Li2015}. A major experimental breakthrough has been reported recently with the observation of superconductivity with a critical temperature above 200 K in hydrogen sulfide (H$_2$S) compressed to $\sim 200$ GPa \cite{Drozdov2015}. Isotopic, magnetic and Meissner-effect measurements have shown that superconductivity is driven by electron-phonon interactions. Electronic structure and electron-phonon coupling calculations \cite{Bernstein2015,Duan2015,Errea2015} and x-ray diffraction experiment \cite{Einaga2016} have determined that the superconducting phase consists of the decomposed product H$_3$S with a cubic structure. The surprisingly high observed $T_c$ raises the possibility that even higher transition temperatures may be attainable in hydrides. So far, the predictions of superconductivity and $T_c$ in hydrides have entirely relied on calculation for selected structures. It is desirable that general rules can be established to understand the underlying mechanisms of high-$T_c$ superconductivity in hydrogen-rich materials. 

In this work, by solving the Eliashberg equations \cite{Eliashberg1960,Carbotte1990,Marsiglio2008}, we analyse the functional derivative of $T_c$ with respect to the electron-phonon spectral function $\alpha^2F(\omega)$ of several representative hydride systems to characterise the most effective vibrational modes for enhancement of superconductivity. Our goal is to develop a strategy for synthesising new compounds with high critical temperatures. The functional derivative $\delta T_c/\delta \alpha^2F(\omega)$ \cite{Bergmann1973} enables us to identify the frequency regions where phonons are most effective in raising $T_c$ \cite{Mitrovic1981,Yao2009,Nicol2015}. We evaluate the functional derivative from the electron-phonon spectral function calculated from linear response theory and density functional perturbation theory, either performed for the present study or taken from reports of previous theoretical studies.

\begin{figure}[ht]
  \centering
  \resizebox{!}{!}{\includegraphics[width=\columnwidth]{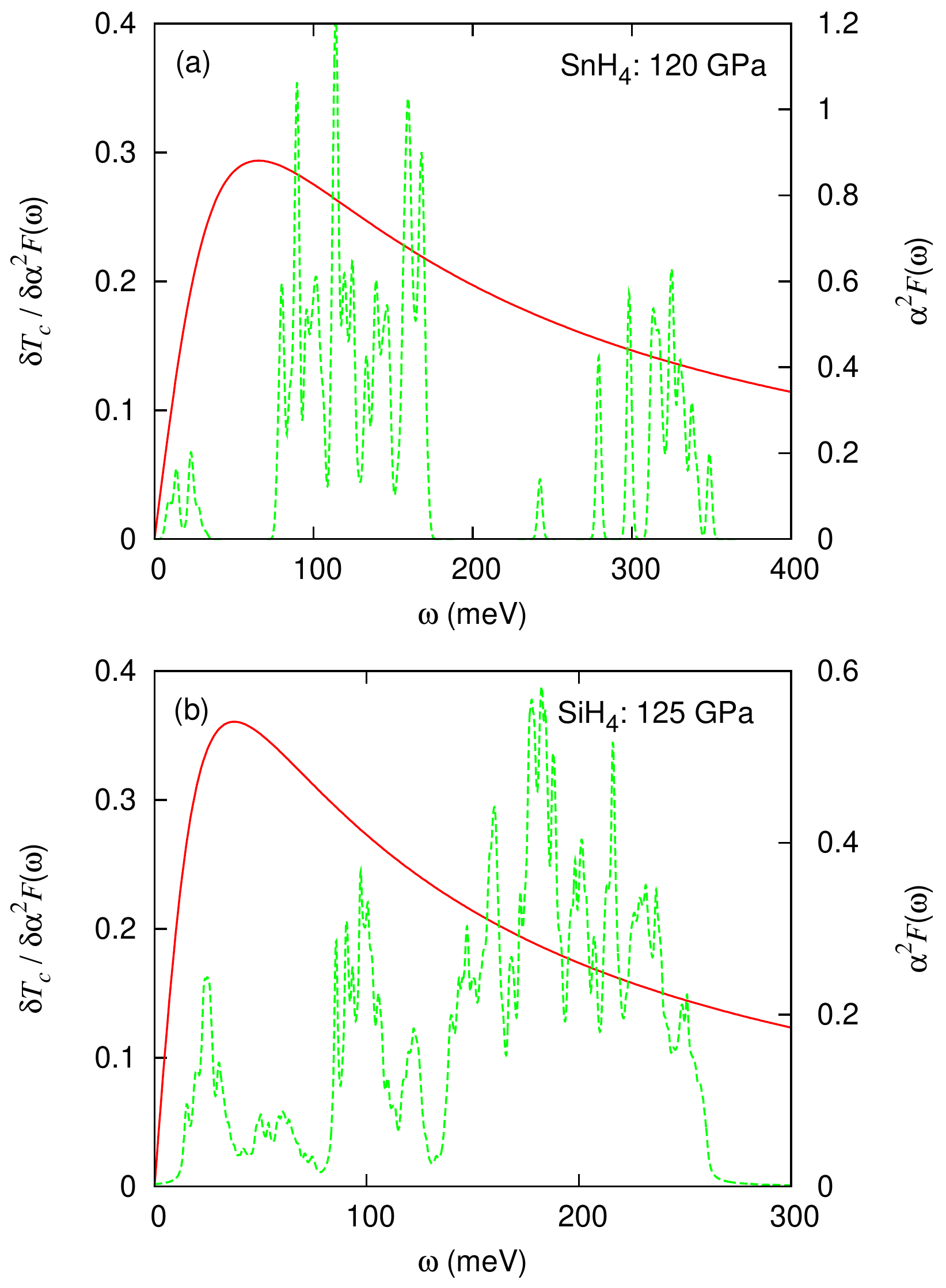}}
  \caption{\label{fig:SnH4_SiH4} (Colour online) The functional derivative $\delta T_c/\delta \alpha^2F(\omega)$ (red solid curve) and the electron-phonon spectral function $\alpha^2F(\omega)$ (green dashed curve) as a function of frequency $\omega$ for (a) SnH$_4$ at 120 GPa and (b) SiH$_4$ at 125 GPa.
  }
\end{figure}

A survey of theoretically predicted hydrogen-dominant main group metallic alloys under high pressures reveals that most alloys adopt a layered structure. For this group of compounds, the calculated $T_c < 100$ K. We have chosen to examine SnH$_4$ in detail as the superconducting metallic phase that is stable between 70 and 160 GPa has a novel layered structure intercalated by ``H$_2$'' units with high $T_c$ \cite{Tse2007}. A remarkable feature of this structure is that the phonon band structure and the spectral function can be separated into three distinct frequency regions, corresponding to lattice (L), Sn-H, and H$_2$ vibrations with increasing frequency, as can be seen in Fig.~\ref{fig:SnH4_SiH4}(a). This unique property facilitates the analysis of the contributions of different vibrational bands to superconductivity, as the spectral function $\alpha^2F(\omega)$ can be decomposed as $\alpha^2F(\omega)_{\rm L}+\alpha^2F(\omega)_{\text{Sn-H}}+\alpha^2F(\omega)_{\text{H-H}}$. The electron-phonon coupling (EPC) parameter $\lambda$ is twice the first inverse moment of the spectral function, $\lambda = 2\int d\omega\,\frac{\alpha^2F(\omega)}{\omega}$. The EPC parameter $\lambda(\omega)$ integrated up to frequency $\omega$ of SnH$_4$ has shown that the lattice and Sn-H librations contribute most to the process of electron-phonon coupling, while there is only very minor contribution from the H-H vibrations \cite{Tse2007}. We have calculated $T_c$ from the individual vibrational bands by solving the Eliashberg equations at 120 GPa, using the Coulomb pseudopotential $\mu^*(\omega_{\text{max}})$ scaled to $\omega_{\text{max}}$, where $\omega_{\text{max}}$ is six times the maximum phonon frequency.
For $\mu^*(\omega_{\text{max}})=0.1$, the entire spectrum yields $T_c\simeq 98$ K. Removing the low-frequency lattice vibrations from the spectrum only reduces $T_c$ roughly by 4 K, while eliminating the high-frequency H-H contribution results in $T_c\simeq 72$ K. In fact, the Sn-H vibrational band by itself yields  $T_c\simeq 66$ K. Thus, the Sn-H vibrations are the most dominant contribution to $T_c$. The functional derivative shown in Fig.~\ref{fig:SnH4_SiH4}(a) is maximum at about 65 meV. This optimal frequency, $\omega_{\text{opt}}$, is close to the onset of the Sn-H bending vibrations. Therefore, it is plausible that variation of the bending vibrations can affect the critical temperature significantly. The optimal frequency is known to be related to the critical temperature by $\omega_{\text{opt}}\sim 7k_BT_c$ \cite{Carbotte1987,Carbotte1990}, where $k_B$ is the Boltzmann constant. Using this relation, $T_c$ is estimated to be roughly 108 K, in good agreement with 98 K from solving the Eliashberg equations.

The functional derivative for the monoclinic SiH$_4$ structure at 125 GPa is presented in Fig.~\ref{fig:SnH4_SiH4}(b). For this compound, no H-H species are present and partition of the phonon spectrum to lattice, Si-H bent (libration) and stretch vibrations is less distinctive. For $\mu^*(\omega_{\text{max}})=0.1$, the optimal vibrational frequency is found to be 38 meV, in the frequency range between the lattice and low-frequency Si-H bending vibrations. $T_c$ estimated from $\omega_{\text{opt}}\sim 7k_BT_c$ is about 63 K, again comparable to 53 K obtained from the Eliashberg equations. The above results on SnH$_4$ and SiH$_4$ highlight the importance of lattice and bending vibrations on the electron-phonon coupling process. It is surprising, however, that the low-frequency lattice vibrations in SnH$_4$ contribute substantially to the EPC parameter \cite{Tse2007}, but not so much to $T_c$. 
This implies that a higher critical temperature cannot necessarily be achieved by simply increasing the mass of the heavier element in hydrides.

\begin{figure*}[t]
  \begin{center}
    \begin{tabular}{p{ 2 \columnwidth}}
      \hspace{-0.5cm}
      \resizebox{!}{!}{\includegraphics[width=1\columnwidth]{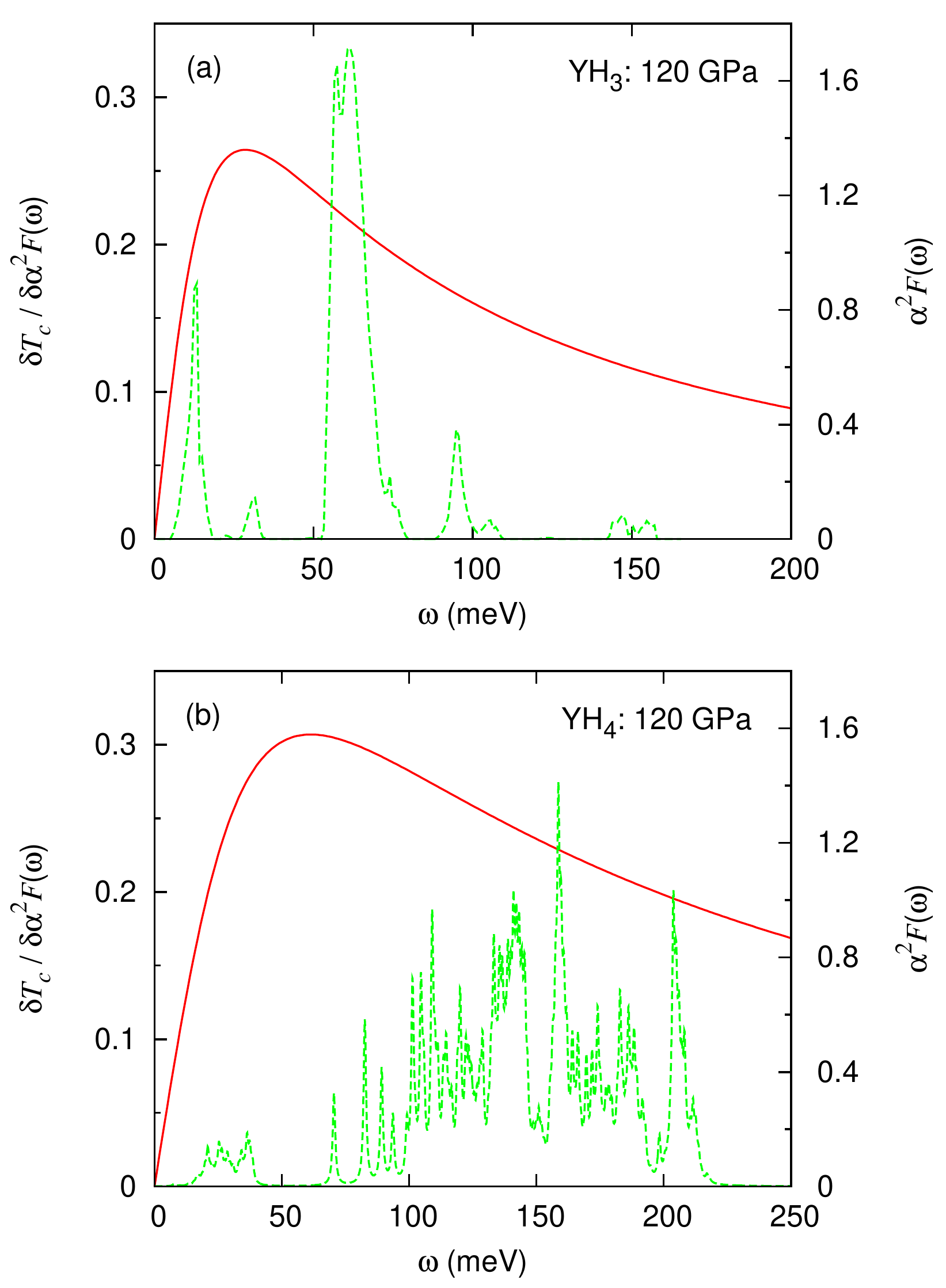}} 
      \hspace{0.5cm}
      \resizebox{!}{!}{\includegraphics[width=1\columnwidth]{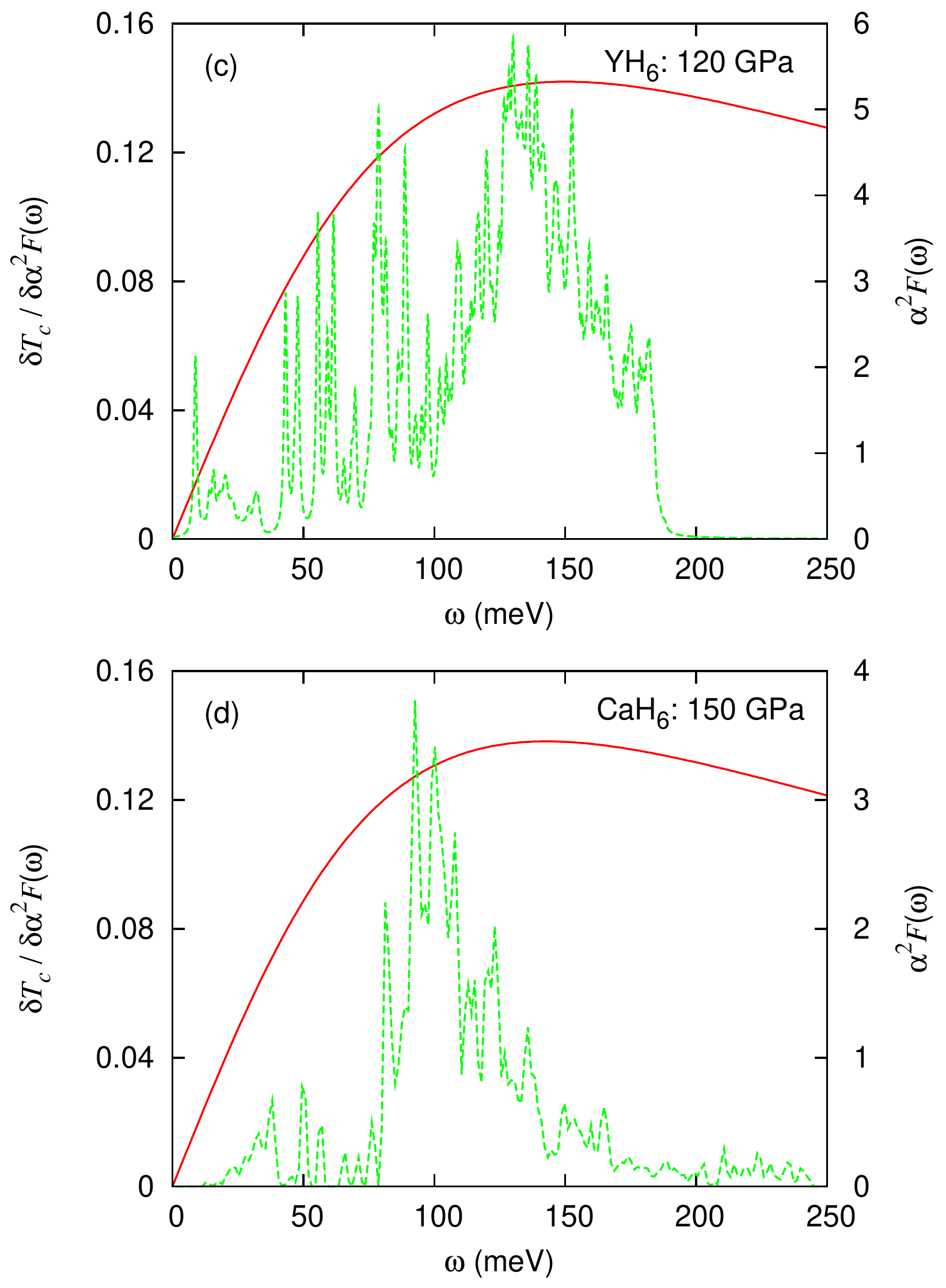}} 
    \end{tabular}
  \end{center}
  \caption{\label{fig:YH_CaH} (Colour online) The functional derivative $\delta T_c/\delta \alpha^2F(\omega)$ (red solid curve) and the electron-phonon spectral function $\alpha^2F(\omega)$ (green dashed curve) as a function of frequency $\omega$ for (a) YH$_3$, (b) YH$_4$ and (c) YH$_6$ at 120 GPa, and (d) CaH$_6$ at 150 GPa.
 }
\end{figure*}

Study of a series of high-pressure yttrium hydrides (YH$_n$, $n = 3$, 4 and 6) offers useful insight into the evolution of crystal structure and the superconducting properties as the hydrogen concentration is increased beyond what is required to satisfy the normal covalency (i.e., YH$_3$). YH$_3$ has been predicted to have a face-centered cubic structure formed from monoatomic H situated in the tetrahedral and octahedral interstitial sites and separated by long distances \cite{Li2015}. This structure is stable from 17.7 GPa to 140 GPa.  Similarly to SnH$_4$, the spectral function can clearly be separated into regions for lattice, Y-H and H-H vibrations. At 17.7 GPa, the H-H vibration energy of 165 meV is much lower than that of normal H$_2$ at the same pressure. The compound is predicted to be superconducting with maximum $T_c$ of 40 K at 17.7 GPa. Upon compression, $T_c$ decreases and superconductivity vanishes between 35 and 44 GPa, only to reappear at a lower value of around 6 K at higher pressure. Two energetically competing hydrogen-rich polymorphs YH$_4$ (YH$_3$+H$_2$) and YH$_6$ (2YH$_3$ + 3H$_2$) have been predicted to be thermodynamically more stable than the product of YH$_3$ and solid H$_2$ above 140 GPa. Both are superconductors with maximum $T_c$ of 85 K and 235 K for YH$_4$ and YH$_6$, respectively. As mentioned above, YH$_3$ has a cubic structure composed of atomic hydrogens. YH$_4$ has tetragonal space group and consists of atomic H and molecular ``H$_2$''. YH$_6$ has a novel cubic cage structure with Y located in the sodalite cages formed by H atoms. The same structure has also been found in CaH$_6$ which too is predicted to be a good superconductor with $T_c$ of 205 K at 150 GPa \cite{Wang2012}. Changes in the H network topology in these structures can be rationalised by a charge transfer model proposed earlier for Ca and Sr hydrides \cite{Wang2012,Wang2015}. Y is trivalent with a valence shell electron configuration $4d^15s^2$. Since $4d$ and $5s$ orbitals are shielded by their respective core orbitals, the valence electrons can be removed easily. Assuming full charge transfer as expected in YH$_3$, the effective electron number (EEN) of each hydrogen is $-1e$ and thus leads to the formation of monatomic hydrogens in the crystal structure. On the other hand, in YH$_4$, the EEN is $(3/4)e$ and hence some of the ``molecular H$_2$'' are preserved, albeit with a long H-H bond length of 1.33 $\angstrom$ at 120 GPa. In YH$_6$, the EEN is further reduced to $(1/2)e$, but to maintain maximum overlaps of the H orbitals, the cage structure is preferred. The features of H-H interactions in these crystal structures suggest the existence of high-frequency H-H vibrons in YH$_4$, YH$_6$, and YH$_3$ with frequency in descending order. This trend has indeed been confirmed by phonon band structure calculations \cite{Li2015}. The predicted maximum $T_c$, however, does not follow this sequence. Therefore, the mean vibrational frequency, $\langle \omega \rangle$, often used in estimating the critical temperature in terms of the McMillan \cite{Mcmillan68} or Allen-Dynes \cite{Allen75prb,Allen75} equation, is not necessarily the only factor to be considered for raising $T_c$.

To gain insight into the role of phonons in the superconducting state, the functional derivative $\delta T_c/\delta \alpha^2F(\omega)$ has been computed for YH$_3$, YH$_4$, and YH$_6$ for $\mu^*(\omega_{\text{max}})=0.1$ and the results are compared in Fig.~\ref{fig:YH_CaH}. The respective optimal frequencies are 29, 61 and 150 meV with $T_c\sim \omega_{\text{opt}}/7k_B$ of 48, 101 and 249 K, respectively, which compare well with $T_c$ of 43, 92 and 247 K calculated from the Eliashberg equations. An interesting aspect of the optimal vibrations is that in YH$_3$ it is maximised at the lattice acoustic translational branch, while in YH$_4$, it originates from the soft vibrational branch of the localised ``molecular H$_2$'' units. In YH$_6$, there is no clear distinction between stretch and bent modes as the H atoms form a 3D connected open sodalite framework. Coupling of these vibrations results in the continuous spectral distribution. These vibrations all participate strongly in the electron-phonon interaction and shift the optimal frequency to higher energy, yielding a higher $T_c$. Moreover, the functional derivative curve is broad and does not taper off as rapidly at higher frequencies as in YH$_3$ and YH$_4$. For comparison, the functional derivative of the isostructural CaH$_6$ is also examined [Fig.~\ref{fig:YH_CaH}(d)]. Once again a continuous distribution of H-dominated vibrations is observed. In this case, the maximum of the functional derivative is located at 143 meV. Although the functional derivative profiles for YH$_6$ and CaH$_6$ are broadly similar, it is important to note that the H-H distance in YH$_6$ of 1.31 $\angstrom$ is significantly larger than 1.24 $\angstrom$ in CaH$_6$ under similar pressure. A larger H-H distance in YH$_6$ can be understood as due to the fact that there is one more valence electron provided by Y than by Ca (three vs. two). Thus, the EEN of the H atom is higher in YH$_6$, leading to a weaker and longer ``molecular H$_2$'' bond. The cutoff frequency for the H-H vibrons in CaH$_6$ (245 meV) is therefore higher than in YH$_6$ ($\sim 189$ meV), but the optimal frequency is lower. 

\begin{figure}[ht]
  \includegraphics[width=\columnwidth]{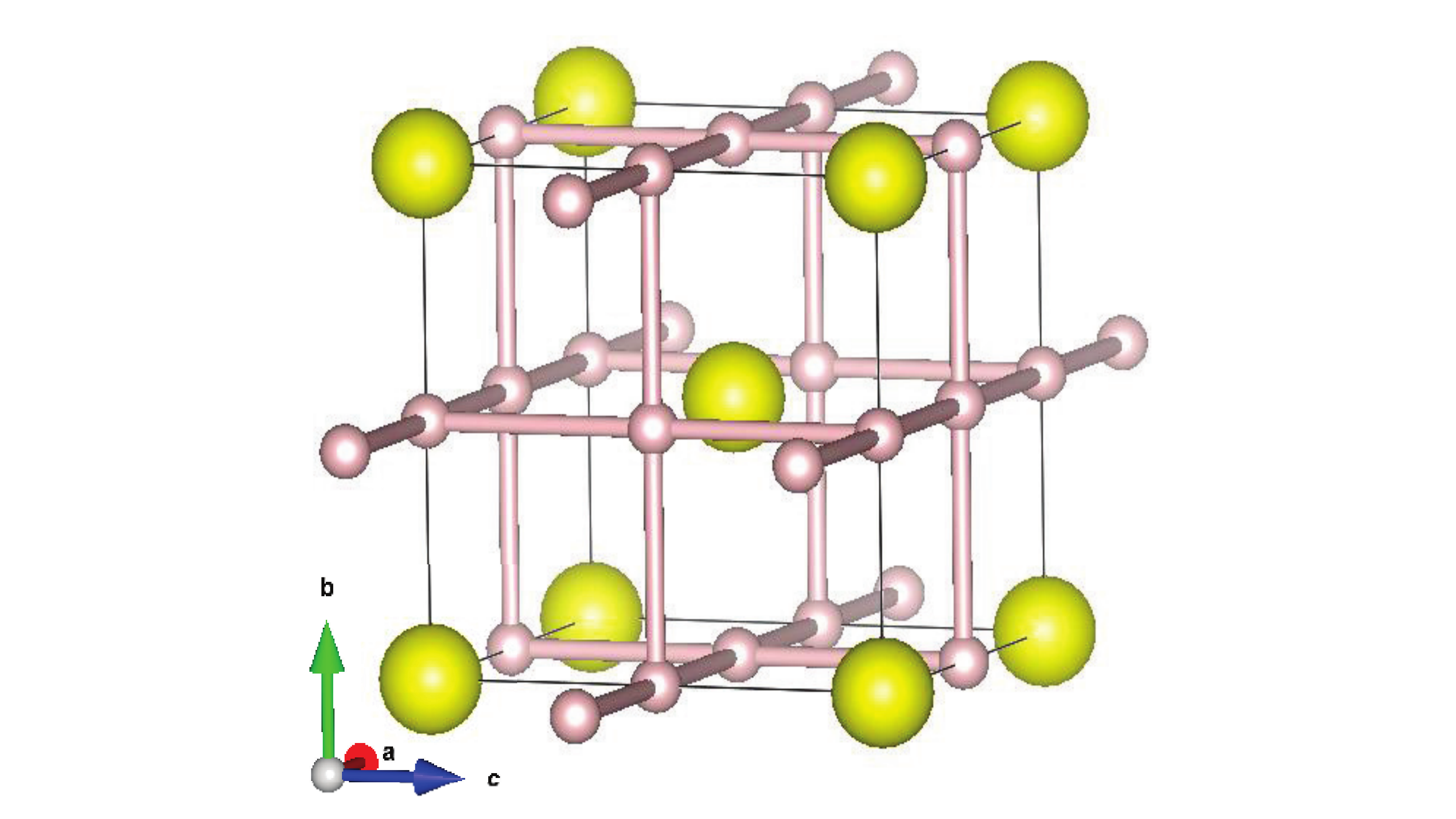}
  \caption{\label{fig:H2S} (Colour online) Crystal structure of H$_3$S.
  }
\end{figure}

The above observation clearly shows that the signature of a high-$T_c$ hydrogen-rich material is the presence of strongly coupled hydrogen-dominant libration and stretch vibrations. As demonstrated above, weak H-H interactions with bond length of 1.2-1.3 $\angstrom$ is desirable. At this bond separation, there is no longer clear distinction between stretch and bent vibrations and all H vibrations participate effectively in the electron-phonon coupling process. Another example is the recently discovered high critical temperature in compressed H$_2$S at $\sim 200$ GPa \cite{Drozdov2015}. The functional derivative of $T_c$ with respect to the spectral function of the candidate H$_3$S has been analysed \cite{Nicol2015} and the major feature in the spectral function is again found to be the hydrogen libration and stretch vibrations being strongly mixed. However, an important point is that the body centered cubic structure may be viewed as the sulfur atoms being enclathrated at the center of a cubic box created by a 3D hydrogen network (Fig.~\ref{fig:H2S}). Therefore, based on the proposed charge transfer model \cite{Wang2012,Wang2015}, the design of materials possessing this unique structural property requires consideration of both the hydrogen concentration and the nature and number of available valence electrons from the donor atom. For example, the sodalite structure is severely distorted when Ca is replaced by the similarly divalent Sr in SrH$_6$, as Sr can no longer be accommodated in the cage due to the large atomic size.

Our findings are consistent with a very recent study of La-H and Y-H systems in terms of density functional theory, where LaH$_{10}$ and YH$_{10}$ have been found to adopt a sodalite-like face-centered cubic structure and have $T_c$ in the range of room temperature \cite{Liu2017}. We have also solved the Eliashberg equations for these systems and have found that $\delta T_c/\delta \alpha^2F(\omega)$ has a broad distribution and decays rather slowly beyond $\omega_{\rm opt}$, similarly to that for YH$_6$ and CaH$_6$ shown in Fig.~\ref{fig:YH_CaH}(c) and (d). In Table~\ref{table} we summarise key quantities for superconductivity for the systems studied in this work, including YH$_{10}$ and LaH$_{10}$ \cite{note}. It can be seen that $\omega_{\rm opt}$ and hence $T_c$ are the highest for YH$_{10}$ at 250 GPa, even though $\lambda$ is smaller compared to YH$_6$ or SrH$_{10}$ (see below). 

\begin{figure*}[ht]
  \begin{center}
    \begin{tabular}{p{ 2 \columnwidth}}
      \hspace{-.1cm}(a)\hspace{\columnwidth}\hspace{.1cm}(c)\\
      \hspace{-.2cm}\resizebox{!}{!}{\includegraphics[width=\columnwidth]{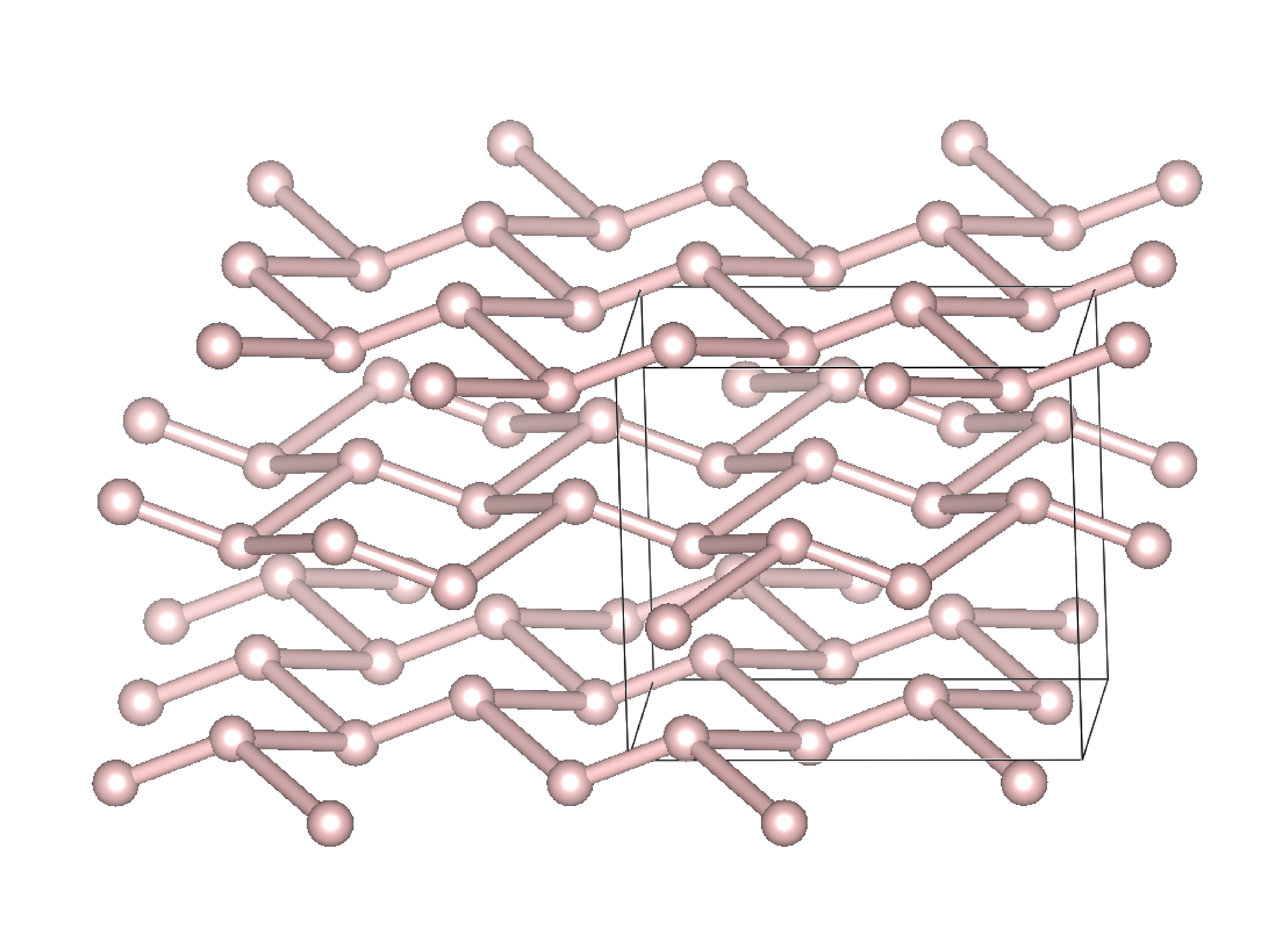}} 
      \hspace{.4cm}
      \resizebox{!}{!}{\includegraphics[width=\columnwidth]{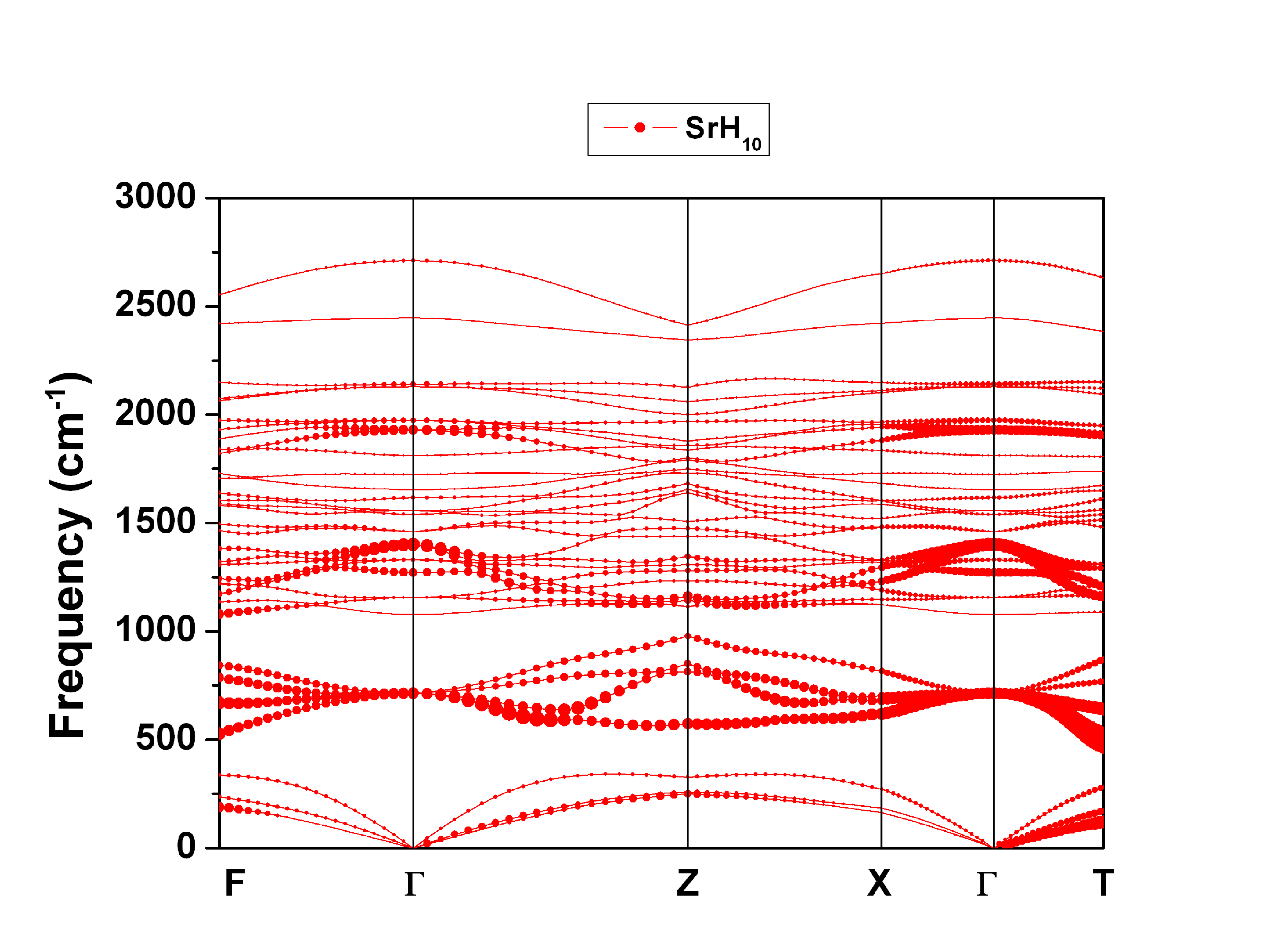}}\\
      \hspace{-.1cm}(b)\hspace{\columnwidth}\hspace{.1cm}\\
      \hspace{-.2cm}\resizebox{!}{!}{\includegraphics[width=\columnwidth]{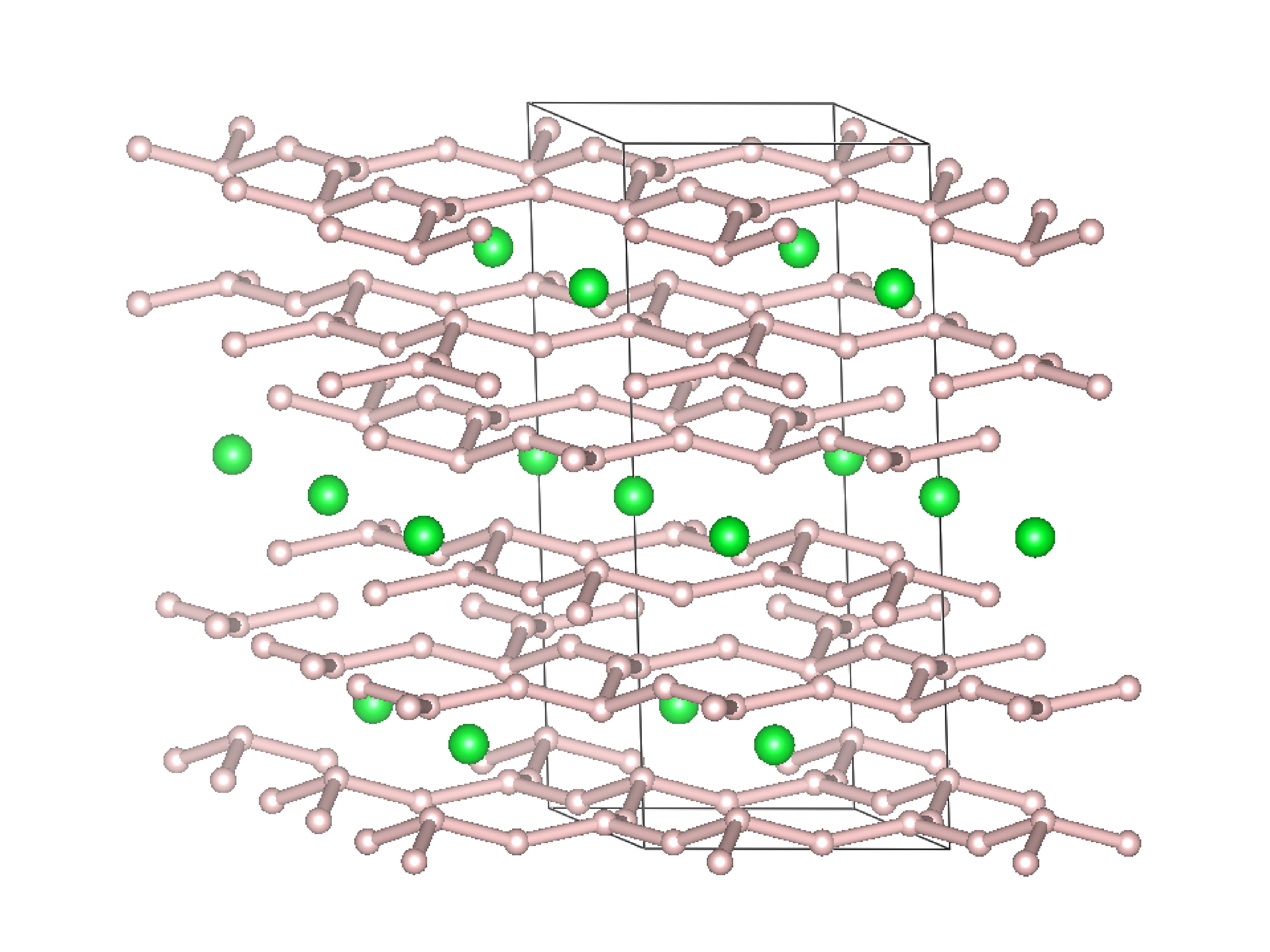}} 
      \hspace{.4cm}
      \resizebox{!}{!}{\includegraphics[width=\columnwidth]{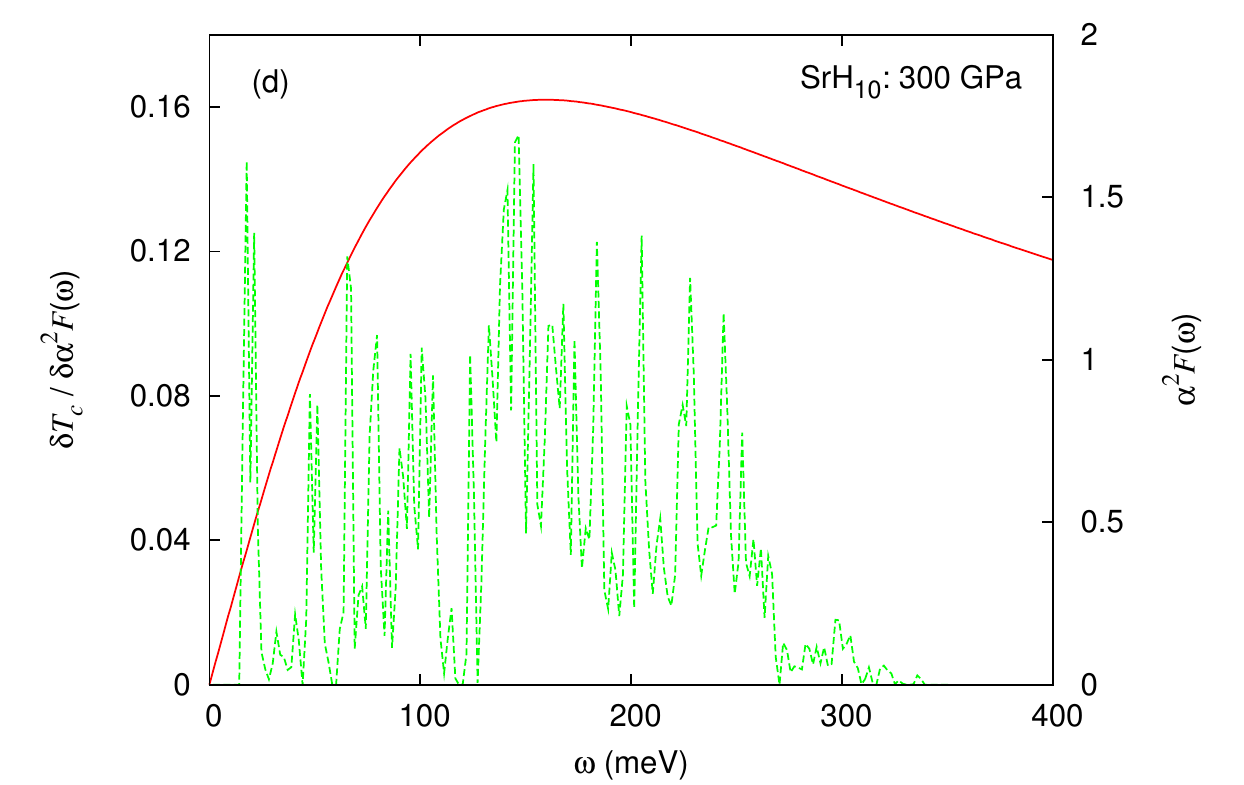}}
    \end{tabular}
  \end{center}
  \caption{\label{fig:SrH10} (Colour online) (a) \emph{Cmca} structure of molecular hydrogen at 300 GPa, (b) rhombohedral crystal structure of SrH$_{10}$, (c) phonon band structure of SrH$_{10}$ at 300 GPa, and (d) $\delta T_c/\delta \alpha^2F(\omega)$ (red solid curve) and $\alpha^2F(\omega)$ (green dashed curve) as a function of frequency $\omega$ for SrH$_{10}$ at 300 GPa.
  }
\end{figure*}

The sodalite cage structure is not the only structural motif that can support strong electron-phonon coupling. Dense molecular hydrogen with the orthorhombic structure (\emph{Cmca}) has been predicted from superconductivity density functional theory to have a critical temperature of 242K at 450 GPa \cite{Cudazzo2008}. The crystal structure at 300 GPa shown in Fig.~\ref{fig:SrH10}(a) is composed of staggered 2D puckered honeycomb layers with interatomic distances alternating between 0.78 and 1.10 $\angstrom$. The closest H-H distance between two layers is 1.27 $\angstrom$. The calculated electron-phonon spectral function is almost continuous up to 494 meV with lattice, libration and molecular vibrations all strongly coupled to the electrons. One may ask, is it possible to construct a similar structural morphology in hydrogen-rich alloys? In a survey of hydrogen-rich strontium hydrides, a high-pressure polymorph, SrH$_{10}$, a rhombohedral crystal with planes of Sr sandwiched between every two puckered honeycomb H layers and H-H bonds alternating between 0.998 and 1.011 $\angstrom$ has been found to be stable above 300 GPa [Fig.~\ref{fig:SrH10}(b)]. The similarity in the structure to the \emph{Cmca} metallic phase of solid hydrogen is striking and suggests potential superconductivity with a high transition temperature. To examine this possibility, the electronic and phonon band structure and electron-phonon coupling for the 300 GPa structure have been calculated using density functional theory. The phonon band structure presented in Fig.~\ref{fig:SrH10}(c) exhibits strong electron-phonon coupling from the librational phonon branches between 50 and 160 meV along the $X\rightarrow \Gamma\rightarrow T$ symmetry direction. SrH$_{10}$ is indeed a superconductor and the calculated isotropic EPC parameter $\lambda$ is 3.08 and the $T_c$ calculated from the Eliashberg equations is 259 K with $\mu^*(\omega_{\text{max}})=0.1$. 
The functional derivative is presented in Fig.~\ref{fig:SrH10}(d). As in metallic hydrogen, the distribution of the phonon modes is almost continuous with the functional derivative maximised at 159 meV. In this case, the high-frequency H-H stretch vibrations centered around 270 meV contribute very little to the overall coupling with electrons. These results confirm the expectation and show that as the sodalite structure, the unique layer H-network morphology is relevant to high-temperature superconductivity. So far, the sodalite and puckered honeycomb layer H-networks are the only two structural features found in hydrogen-rich materials possessing very high $T_c$ ($> 200$ K) by theoretical calculations.

\begin{table*}
  \begin{center}
    \caption{\label{table} $T_c$ obtained by solving the Eliashberg equations, the EPC parameter $\lambda$, the optimal frequency $\omega_{\text{opt}}$, the average phonon frequency $\langle \omega \rangle$, $\omega_{\rm log}$ in the Allen-Dynes-modified McMillan equation \cite{Allen75}, and the H-H distance in the H network for various systems studied. In SiH$_4$ hydrogens do not form a network by themselves. $\mu^*(\omega_{\text{max}})=0.1$ has been used throughout.} 
    \begin{tabular}{lllllllll}
      \hline \hline 
      \vspace{-0.3cm}\\
      System \hspace{.1cm} & Pressure (GPa) \hspace{.2cm} & $T_c$ (K) & \hspace{.3cm} $\lambda$ \hspace{.5cm} & $\omega_{\text{opt}}$ (meV) \hspace{.2cm} & $\omega_{\text{opt}}/7k_B$ (K) \hspace{.2cm} & $\langle \omega \rangle$ (meV) \hspace{.2cm} & $\omega_{\rm log}$ (meV) \hspace{.1cm} & H-H distance ($\angstrom$) \\ 
      \hline
      \vspace{-0.3cm}\\
      SiH$_4$   & \hspace{.2cm} 125 &\hspace{0.05cm} 53 &\hspace{.1cm}  0.89 & \hspace{.27cm} 38 & \hspace{.37cm} 63 & 111 & \hspace{.15cm} 79 &\hspace{.6cm} -- \\
      SnH$_4$   & \hspace{.2cm} 120 &\hspace{0.05cm} 98 &\hspace{.1cm}  1.20 & \hspace{.27cm} 65 &\hspace{.2cm} 108 & 111 & \hspace{.15cm} 76 &\hspace{.3cm} 0.841 \\
      YH$_6$    & \hspace{.2cm} 120 & 247 &\hspace{.1cm}  3.19 &\hspace{.1cm} 150 &\hspace{.2cm} 249 & \hspace{.05cm} 87 & \hspace{.15cm} 63  &\hspace{.3cm} 1.306 \\ 
      YH$_{10}$  & \hspace{.2cm} 250 & 291 &\hspace{.1cm}  2.67 &\hspace{.1cm} 177 &\hspace{.2cm} 293 & \hspace{.05cm} 97 & \hspace{.15cm} 95 &\hspace{.3cm} 1.132 \\ 
      YH$_{10}$  & \hspace{.2cm} 300 & 275 &\hspace{.1cm}  2.00 &\hspace{.1cm} 170 &\hspace{.2cm} 282 & 147 &\hspace{.0cm} 125 &\hspace{.3cm} 1.029 \\
      LaH$_{10}$ & \hspace{.2cm} 300 & 231 &\hspace{.1cm}  1.74 &\hspace{.1cm} 144 &\hspace{.2cm} 239 & 142 &\hspace{.0cm} 123 &\hspace{.3cm} 1.076 \\ 
      CaH$_6$   & \hspace{.2cm} 150 & 235 &\hspace{.1cm}  2.71 &\hspace{.1cm} 143 &\hspace{.2cm} 237 & \hspace{.05cm} 96 & \hspace{.15cm} 87 &\hspace{.3cm} 1.238 \\
      SrH$_{10}$ & \hspace{.2cm} 300 & 259 &\hspace{.1cm}  3.08 &\hspace{.1cm} 159 &\hspace{.2cm} 264 &\hspace{.05cm} 96 & \hspace{.15cm} 66 &\hspace{.3cm}  0.997 \\
      \hline \hline
    \end{tabular}
  \end{center}
\end{table*}

In summary, we have shown from analysis of the structures and functional derivative of selected high-pressure hydrides that the metallic-H bent (or librational) vibrations are most effective in enhancing the superconducting transition temperature. Furthermore, since the vibration profile (or the electron-phonon spectral function) is intimately related to the crystal structure, two types of H networks, the sodalite and puckered honeycomb layer structures with strong mixing of stretch and bent vibrations are most likely to lead to strong electron-phonon coupling for all the modes. Can room-temperature superconductivity be achieved if the phonon-mediated Eliashberg theory is valid? From the relation $\omega_{\text{opt}}\sim 7k_BT_c$, the optimal frequency should be 180 meV (1450 cm$^{-1}$) for a critical temperature $T_c = 300$ K. Among the results presented in Table~\ref{table}, YH$_{10}$ at 250 GPa is close to ideal. The key is to prepare a system with a broad vibration distribution and efficient electron-phonon coupling close to this frequency. As illustrated above, the functional derivative of the sodalite structures (CaH$_6$ and YH$_6$) does not decrease as quickly above the optimal frequency as the other structures, indicating that all the modes, with the exception of lattice vibrations, are very efficient in enhancing $T_c$. Moreover, the spectral function shows that the electron-phonon coupling strength (the area under the spectrum) is uniformly large for these structures. In principle, it is plausible to design materials with such characteristics, perhaps by choosing a di- or trivalent element with a valence electron ionization energy (electron donating property) intermediate between Ca and Y such that the H-H distance in the sodalite structure is about halfway ($\simeq 1.27 \angstrom$). This will decrease the stretch frequency but maintain strong electron-phonon coupling. Another possibility is to increase the H$_2$ concentration in compounds with electron donating atoms. As observed in SrH$_{10}$ above, the metallic atoms help to reduce the pressure required to form the puckered ``molecular'' H$_2$ layers in superconducting solid hydrogen.

The research was supported by the Natural Sciences and Engineering Research Council of Canada and the Canada Foundation for Innovation. H. L. acknowledges support by EFree, an Energy Frontier Research Center funded by the DOE, Office of Science, Basic Energy Sciences under Award No. DE-SC-0001057.

\clearpage
\newpage
\widetext
\onecolumngrid

\end{document}